\documentclass[aps,prl,twocolumn,groupedaddress,showpacs,preprintnumbers
]{revtex4}
\usepackage{graphicx}
\usepackage{bm}

\begin{document}

\title{Duplex-single strand denaturing transition in DNA oligomers}

\author{G. Zocchi, A. Omerzu, T. Kuriabova, J. Rudnick and G.
Gr\"{u}ner}

\affiliation{Department of Physics and Astronomy
University of California Los Angeles
Los Angeles CA 90095
}

\begin{abstract}
We have measured the temperature driven denaturing, or melting
transition in poly d(A)-poly d(T) DNA oligomers of various lengths in
different buffer conditions.  Our findings are in clear disagreement
with two state, reaction kinetics model, and we find that the
so-called zipper model, where denaturing proceeds through opening of
the duplex at the ends describes well the temperature dependence of
the average number of open base pairs.  Analysis of the length
dependence of the transition parameters however suggest that bubble
formation is important and that the transition, in the thermodynamic
limit, is continuous, albeit close to first order.
\end{abstract}


\maketitle

Replication and translation, the fundamental processes in biology,
involve the separation of double stranded DNA \cite{watson}. 
Duplex-single strand transition occurs also when DNA is
heated\cite{wartell,azbel}, upon changing the buffer surrounding the
DNA, or under the influence of an external force or
torque\cite{noy,lubensky,cocco}.  This so-called denaturing transition
has been extensively studied both in natural and synthetic duplexes. 
Aside from its inherent interest, the process is also of primary
importance for applications in the biotech arena.  For example,
thermal denaturation is an essential step in the Polymerase Chain
Reaction amplification procedure.

For the thermally driven transition of native DNA segments of several
thousand base pairs the transition occurs in discrete steps, these
steps being determined by the particular base pair sequences
\cite{marx,turner}.  In long native DNA the transition is smooth,
presumably reflecting a large number of discrete steps occurring at
different temperatures.  The denaturing transition was also
investigated in short oligomers, in the vast majority of cases with
random distribution of CG and TA base pairs.

Several models have been proposed to describe this duplex-single
strand transition.  The standard view of the thermal denaturation of
DNA is that it represents the classic competition between energy and
entropy.  At low temperatures, the thermodynamics is dominated by the
binding energy of the base pairs.  As the temperature is raised,
sections of the DNA separate to take advantage of the greater entropy
available to two separated single DNA strands, as opposed to smaller
entropy to be found in the smaller configuration space accessible to
tightly inter-wound double-stranded DNA. In the thermodynamic limit
(or for circular DNA) melting transition occurs because of the growth
and accretion of denaturation bubbles.  For finite size oligomers
``fraying'' at the ends of a section of linear DNA is likely to be
important as we will discuss below.

Here we address the simplest possible scenario: the denaturing of
short oligomers, where each strand is a homopolymer (composed of
identical base pairs).  Under such circumstances variations of base
pair interaction energies (different for CG and AT pairs) do not
occur, and the oligomer can be regarded as a duplex held together by
identical base pair interaction energy at the different sites.  We
are not aware of experiments which address the situation where
differences between binding energies associated with different base
pairs and other complications do not arise and which thus would allow
the experimental test of simple, but important descriptions of the
melting transition.
 
For finite oligomers, the following argument can be made: the binding
energy between two bases located at the end of the molecule is smaller
than the binding energy for pairs away from the ends, consequently the
unbinding occurs most likely by a ``zipper'' like sequential opening
of the base pairs, starting at the ends where the binding is weakest. 
Such model for DNA melting has been proposed by C. Kittel
\cite{kittel}.  In this so-called zipper model, the melting of a
linear DNA oligomer occurs entirely as a result of strand separation
at the ends.  One assigns an energy $-\epsilon_{0}$ to each bound base
pair, and an entropy equal to $S_{0}$ to each unbound pair.  Then, the
partition function of an N-base-pair oligomer is given by
\begin{equation}
 Q_{N}=\sum_{N_{1}+N_{2} \le N}e^{(N_{1}+N_{2})(S_{0}- 
 \epsilon_{0})/k_{B}T}
 \label{eq:1}
 \end{equation}
where $N_{1}$ and $N_{2}$ are the number of separated base pairs at
the two ends of the linear strand.  The zipper model, and this
partition function, ought to be reasonably accurate as long as one can
ignore the effects of excluded volume, which should be the case for
oligomers that are not too long, and if the oligomer is uniform. 
Here, we use the zipper model as an fitting form for the experimental
data.  A much simpler model, assuming that only two configurations
occur, completely closed and completely separated strands, has also
been used to describe denaturing.  We call this model the
``two-state'' model; for this description the appropriate partition
function has the form:
\begin{equation}
Q_{N}= 1+\exp \left[-N \left(\frac{\epsilon_{0}}{k_{B}T} - S_{0}  
\right)\right]
\label{eq:2}
\end{equation}

The two-state model predicts a first order melting transition in the
thermodynamic limit ($N \rightarrow \infty$).  As it turns out, the
zipper model also leads to the same conclusion5 \cite{kittel}.  Using
the partition functions as given in Eqs.  (\ref{eq:1}) and
(\ref{eq:2}), physical quantities, such as the average number of
paired bases, and the distribution of oligomers with different numbers
of open base pairs, can be calculated.

The zipper model is a theoretical scenario in which denaturation takes
place via unraveling at the ends of a DNA duplex.  However, another
contributing factor in the thermal denaturation of long DNA molecules
is the denatured ``bubble,'' a portion of denatured DNA bounded by
duplexed segments.  Entropic considerations militate in favor of an
accumulation of such bubbles in a sufficiently long DNA molecule.  In
fact, the most physically reasonable picture of the denaturing
transition is in terms of the proliferation and merging of denatured
bubbles.  Poland and Scheraga \cite{ps1,ps2} have proposed a model of
the transition based on this notion.  This model admits of elaboration
and is amenable to analysis in the context of field-theoretical
approaches to the statistical mechanics of critical phenomena.  It is
consistent with either a continuous or a first order transition,
depending on the influence of base pair inhomogeneity
\cite{cule,singh} and excluded volume \cite{kafri}.  Furthermore, the
Poland-Scheraga model, along with the closely-related model of Peyrard
and Bishop \cite{peyrard}, produces results that are consistent with
scaling and hyperscaling analysis of both continuous and first order
transitions \cite{note}.
 
In this paper we focus on the average number of open base pairs as
function of temperature, using the intensity of the UV absorption at
the wavelength of 260 nm.  This parameter predominantly measures base
stacking which is directly related to the number of open base pairs
\cite{cantor1}.  Other spectroscopic methods are also available for
monitoring the melting transition.  In a separate study \cite{gruner}
we have demonstrated that for the oligomer dA15/dT15 three different
spectroscopic methods (UV absorption, CD spectroscopy, and a
fluorescence based method) give rise to identical (within experimental
error) melting curves.  We believe therefore that the assumption we
make, namely that the measured UV absorption correctly represents the
average number of open base pairs is justified.

We have used synthetic poly(A) and poly(T) oligomers of three
different lengths - 15, 30 and 60 bases, PAGE purified, purchased from
Operon Technologies.  Single strands were dissolved in 1.5 M Phosphate
Buffered Saline (PBS).  For recombination, solutions of complementary
strands were mixed in equimolar ratio, warmed up to 90$^{\circ}$C in a
water bath, followed by a slow cool down to room temperature.  This
resulted in complete recombination of the complementary strands as
confirmed with hypochromicity measurements.  A quantity of few $\mu$L
was isolated from the stock solutions and it was dissolved in 500 mL
of 50 mM PBS buffer adjusting the final DNA concentration to 1 OD. For
measurement in buffers of higher molarities the appropriate volumes of
1.5 M PBS were added to the samples in order to obtain 100 mM and 200
mM buffer concentrations.  This led to a slight dilution of DNA
solutions.  Absorption measurements were done in a standard quartz
cuvette with a Bekman-Coulter 640 UV/Vis spectrophotometer with an
integrated Peltier heating block and a temperature controller that
enable temperature control between 10$^{\circ}$C and 90$^{\circ}$C.
Temperature dependent absorption measurements were done in steps of 1
K. Before the absorption measurement the samples were thermalized at
every temperature for 5 min---the time needed for the cuvette and
solution inside to reach the temperature of the heating block.  The
absorption is smaller for a DNA duplex than for the same DNA in single
strand form \cite{cantor1}; this is referred to as hyperchromicity. 
The main component of this effect is the screening of the intra-base
excitations by dipole-dipole interactions between stacked bases, with
significantly smaller screening for a single strand DNA on which bases
are unstacked.  For poly d(A)-poly d(T) this difference, the ratio of
the intensity for single strand and duplex DNA, is 1.4 (see below). 
In Fig.  1, the temperature dependence of the UV absorption intensity
is displayed for three different oligomer lengths.  For all oligomers
we observe a smooth transition from the duplex to the single strand
state, with the transition temperature (defined as the half-point of
the transition---see below) and width, increasing with increasing
length.  The linear slope visible in the melting curves after the
sigmoidal transition region is a well-known phenomenon attributed to
residual base unstacking in the single strands.  The linear slope
before the transition is indicative of temperature driven
conformational changes in the double helix; this phenomenon, known as
``premelting,'' is not well understood \cite{cantor2}.  These
phenomena are not accounted for in the models above: the first one is
not related to strand separation, while the degrees of freedom
relevant for the second are not taken into account by the zipper
model.

We start with a comparison between the experimental results and the
two state and zipper models discussed above.  Such comparison is shown
in Fig.  1.  
\begin{figure}[htb]
\includegraphics[height=2in]{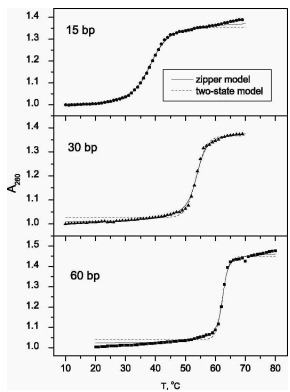}
\caption{Temperature dependence of the UV absorption measured at 260
nm for poly d(A)-poly d(T) oligomers of three different lengths - 15,
30 and 60 base pairs (bp).  All curves are at a molarity of 100mM.}
\label{fig:1}
\end{figure}
In the zipper model, the fitting parameter $\epsilon_{0}$
(in Kelvin) was allowed to vary between 6405 and 7090, and the
parameter $S_{0}$ was fixed at 20.8.  The points utilized for the fit were
those closest to the transition.  The dominant temperature dependence
of the UV absorption in the vicinity of the melting transition is due
to the separation of base pairs.  At temperatures significantly higher
and lower than the nominal melting temperature, the absorbance
exhibits a temperature dependence as a result of effects that are
unrelated to the denaturation of the DNA, as discussed above.

The fitting results for zipper model are summarized in the table below
(Table 1).  The binding energy increases with salt concentration,
because of ionic screening (the temperature at the midpoint is ).The
fact that the two-state model fits the data for the 15mer but not the
60mer indicates that gradual opening of the duplex plays an important
role in the melting in the case of larger oligomers.

\begin{table}
\begin{tabular}{|c|c|c|c|}
\hline 
bp & molarity/mM & $ - \epsilon_{0}/$K & $S_{0}$ \\
\hline
\hline
 & 50 & 6405 & \\
\hline 15 & 100 & 6524 & 20.8 \\
\hline
& 200 & 6611  & \\
\hline
\hline
& 50 & 6742 & \\
\hline
30 & 100 & 6822 & 20.8 \\
\hline
& 200 & 6916 & \\
\hline
\hline
& 50 & 6848 & \\
\hline
60 & 100 & 6991 & 20.8 \\
\hline
& 200 & 7090 & \\
\hline
\end{tabular}

\caption{Parameters used to fit (see text) the measured absorption
curves to the zipper model.  \label{table:1}}
\end{table}

Figure 2 displays the temperature derivative of the UV absorption.  
\begin{figure}[htb]
\includegraphics[height=2in]{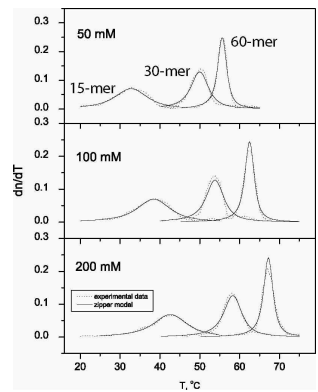}
\caption{Temperature dependence of the derivative of the UV absorption
of poly d(A)-poly d(T) at three different molarities.  Here, the
absorption has been normalized so that the area under each peak is
one.  The full line is obtained fitting the absorption curves with the
zipper model and then taking the derivative (``zipper
interpolation''), the dotted line is obtained using a three-point
interpolation of the experimental data.  The parameters used for the
zipper fits are given in the table.  In all cases, the height of the
maximum, and the temperature at which this maximum occurs, increase
monotonically with the size of the oligomer.  }
\label{fig:2}
\end{figure}
In this case the absorption has been normalized so that the integrated
weight under each peak is equal to one in all cases.  Two curves are
shown for each data set.  One represents the results of taking the
derivative of the best fit of the zipper model to the data.  The other
was obtained by taking the temperature derivative of a three-point
Lagrange interpolation through the data.  While there are systematic
differences between the two derivative curves, the tendencies of both
are the same, as can be seen in Fig.  3, representing a log-log plot of
the maximum of the derivative curve against the number of base pairs
in the oligomers.  This last figure is relevant to the analysis
discussed below.
\begin{figure}[htb]
\includegraphics[height=2.5in]{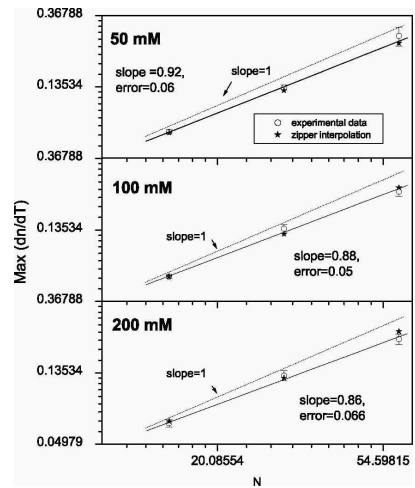}
\caption{Log-log plots of the maximum of the derivative of the
melting curves against the size of the oligomer at the three
molarities.  The two sets of points refer to two different
interpolations of the data, see caption to Fig.2.  The best-fit linear
regression fit is shown.  Also displayed is a line with unit slope,
representing a first order transition in the large N limit.  }
\label{fig:3}
\end{figure}

In light of the likely relevance of standard scaling notions to DNA
melting, we have applied finite size scaling to our data.  According
to finite size scaling analysis, the specific heat of a $d$-dimensional
system with (linear) size $L$ that undergoes a continuous phase
transition will take the form
\begin{equation}
c \propto |t|^{d \nu - 2}f \left(L|t|^{\nu}\right)
\label{eq:3}
\end{equation}
in the immediate vicinity of the transition.  In the above expression,
$t$ is the reduced temperature and $\nu$ is the correlation function
exponent.  As no finite system exhibits thermodynamic singularities,
the behavior of the function $f$ is such that the singularity in $t$
in the prefactor is cancelled as $t \rightarrow 0$.  This implies a
specific heat at the bulk transition temperature that goes as
$L^{(2-d\nu)/\nu}$.  In the case at hand, $d=1$, so the specific heat
at the bulk transition temperature scales as $L^{(2-\nu)/\nu}$.  This
sort of dependence on $L$ also characterizes the maximum in the
specific heat.  The temperature derivative of the UV absorption should
behave in essentially the same way as does the specific heat at the
denaturing transition, against $L$, and we have evaluated the maximum
values of dn/dT using two procedures.  The first involves the Lagrange
interpolation through the data.  The second, which we call the
``Zipper interpolation'' refers to a theoretical fit to the observed
temperature dependence, using Eq.  (\ref{eq:1}), with $\epsilon_{0}$
as a free parameter for each oligomer and identifying the maximum of
the derivative of the fit.  The length dependence of the maxima are
displayed in Fig.  3, and we find that the optimal fit is consistent
with a specific heat that scales as $L^{0.86}$, which implies a $\nu =
1.075$ and a specific heat diverging as $|t|^{-0.925}$ in an infinite
system.  Consequently, the transition is continuous, but close to
first order.  Our findings are, therefore, consistent with the picture
that bubble formation, in addition to opening at the ends, is
important for the denaturing process.  This is so even in the case of
relatively short oligomers.

The experiments and analysis given above lead to several conclusions. 
First, it is clear from Fig 2 that the zipper model is a better fit to
the experiments than the two state model for longer oligomers.  This
is in agreement with observations \cite{cantor1} that the dependence
of the melting temperature on oligomer concentration is not
significant above a length of about 14, while for shorter oligomers
the transition is a chemical equilibrium between single strand and
duplex species, which depends on concentration \cite{cantor3}. 
Contrasting our results obtained on different lengths suggest that
bubble formation, and a scaling scenario of the denaturing transition,
is likely to be important.  Experiments on longer oligomers, where the
bubble formation process is expected to be more important would be
desirable in order to distinguish between the zipper model and phase
transition scenario.  Finally we note that we have analyzed only the
average, or mean number of open base pairs.  The measurements
described here do not offer insight into effects associated with
fluctuations.

\bibliography{melting}

\begin{thebibliography}{20}
\expandafter\ifx\csname natexlab\endcsname\relax\def\natexlab#1{#1}\fi
\expandafter\ifx\csname bibnamefont\endcsname\relax
  \def\bibnamefont#1{#1}\fi
\expandafter\ifx\csname bibfnamefont\endcsname\relax
  \def\bibfnamefont#1{#1}\fi
\expandafter\ifx\csname citenamefont\endcsname\relax
  \def\citenamefont#1{#1}\fi
\expandafter\ifx\csname url\endcsname\relax
  \def\url#1{\texttt{#1}}\fi
\expandafter\ifx\csname urlprefix\endcsname\relax\def\urlprefix{URL }\fi
\providecommand{\bibinfo}[2]{#2}
\providecommand{\eprint}[2][]{\url{#2}}

\bibitem[{\citenamefont{Watson}(1988)}]{watson}
\bibinfo{author}{\bibfnamefont{J.~D.} \bibnamefont{Watson}},
  \emph{\bibinfo{title}{Molecular biology of the gene}}
  (\bibinfo{publisher}{Benjamin/Cummings Pub. Co.}, \bibinfo{address}{Menlo
  Park, Calif.}, \bibinfo{year}{1988}), \bibinfo{edition}{4th} ed.

\bibitem[{\citenamefont{Wartell and Benight}(1985)}]{wartell}
\bibinfo{author}{\bibfnamefont{R.~M.} \bibnamefont{Wartell}} \bibnamefont{and}
  \bibinfo{author}{\bibfnamefont{A.~S.} \bibnamefont{Benight}},
  \bibinfo{journal}{Physics Reports} \textbf{\bibinfo{volume}{126}},
  \bibinfo{pages}{67} (\bibinfo{year}{1985}).

\bibitem[{\citenamefont{Azbel}(1979)}]{azbel}
\bibinfo{author}{\bibfnamefont{M.~Y.} \bibnamefont{Azbel}},
  \bibinfo{journal}{Physical Review A (General Physics)}
  \textbf{\bibinfo{volume}{20}}, \bibinfo{pages}{1671} (\bibinfo{year}{1979}),
  \bibinfo{note}{article}.

\bibitem[{\citenamefont{Noy et~al.}(1997)\citenamefont{Noy, Vezenov, Kayyem,
  Meade, and Lieber}}]{noy}
\bibinfo{author}{\bibfnamefont{A.}~\bibnamefont{Noy}},
  \bibinfo{author}{\bibfnamefont{D.~V.} \bibnamefont{Vezenov}},
  \bibinfo{author}{\bibfnamefont{J.~F.} \bibnamefont{Kayyem}},
  \bibinfo{author}{\bibfnamefont{T.~J.} \bibnamefont{Meade}}, \bibnamefont{and}
  \bibinfo{author}{\bibfnamefont{C.~M.} \bibnamefont{Lieber}},
  \bibinfo{journal}{Chemistry and Biology} \textbf{\bibinfo{volume}{4}},
  \bibinfo{pages}{519} (\bibinfo{year}{1997}).

\bibitem[{\citenamefont{Lubensky and Nelson}(2000)}]{lubensky}
\bibinfo{author}{\bibfnamefont{D.~K.} \bibnamefont{Lubensky}} \bibnamefont{and}
  \bibinfo{author}{\bibfnamefont{D.~R.} \bibnamefont{Nelson}},
  \bibinfo{journal}{Physical Review Letters} \textbf{\bibinfo{volume}{85}},
  \bibinfo{pages}{1572} (\bibinfo{year}{2000}).

\bibitem[{\citenamefont{Cocco and Monasson}(1999)}]{cocco}
\bibinfo{author}{\bibfnamefont{S.}~\bibnamefont{Cocco}} \bibnamefont{and}
  \bibinfo{author}{\bibfnamefont{R.}~\bibnamefont{Monasson}},
  \bibinfo{journal}{Physical Review Letters} \textbf{\bibinfo{volume}{83}},
  \bibinfo{pages}{5178} (\bibinfo{year}{1999}).

\bibitem[{\citenamefont{Marx et~al.}(2000)\citenamefont{Marx, Bizzaro, Blake,
  Hsien~Tsai, and Feng~Tao}}]{marx}
\bibinfo{author}{\bibfnamefont{K.~A.} \bibnamefont{Marx}},
  \bibinfo{author}{\bibfnamefont{J.~W.} \bibnamefont{Bizzaro}},
  \bibinfo{author}{\bibfnamefont{R.~D.} \bibnamefont{Blake}},
  \bibinfo{author}{\bibfnamefont{M.}~\bibnamefont{Hsien~Tsai}},
  \bibnamefont{and} \bibinfo{author}{\bibfnamefont{L.}~\bibnamefont{Feng~Tao}},
  \bibinfo{journal}{Molecular and Biochemical Parasitology}
  \textbf{\bibinfo{volume}{107}}, \bibinfo{pages}{303} (\bibinfo{year}{2000}).

\bibitem[{\citenamefont{Turner et~al.}(1990)\citenamefont{Turner, Ahlquist, and
  White}}]{turner}
\bibinfo{author}{\bibfnamefont{T.~F.} \bibnamefont{Turner}},
  \bibinfo{author}{\bibfnamefont{J.~E.} \bibnamefont{Ahlquist}},
  \bibnamefont{and} \bibinfo{author}{\bibfnamefont{M.~M.} \bibnamefont{White}},
  \bibinfo{journal}{Journal of Fish Biology} \textbf{\bibinfo{volume}{37}},
  \bibinfo{pages}{531} (\bibinfo{year}{1990}), \bibinfo{note}{12-4 FIELD
  Section Title:Nonmammalian Biochemistry Dep. Zool. Biomed. Sci.,Ohio
  Univ.,Athens,OH,USA. FIELD URL: written in English.}

\bibitem[{\citenamefont{Kittel}(1969)}]{kittel}
\bibinfo{author}{\bibfnamefont{C.}~\bibnamefont{Kittel}},
  \bibinfo{journal}{American Journal of Physics} \textbf{\bibinfo{volume}{37}},
  \bibinfo{pages}{917} (\bibinfo{year}{1969}).

\bibitem[{\citenamefont{Poland and Scheraga}(1966{\natexlab{a}})}]{ps1}
\bibinfo{author}{\bibfnamefont{D.}~\bibnamefont{Poland}} \bibnamefont{and}
  \bibinfo{author}{\bibfnamefont{H.~A.} \bibnamefont{Scheraga}},
  \bibinfo{journal}{J. Chem. Phys.} \textbf{\bibinfo{volume}{45}},
  \bibinfo{pages}{1456} (\bibinfo{year}{1966}{\natexlab{a}}).

\bibitem[{\citenamefont{Poland and Scheraga}(1966{\natexlab{b}})}]{ps2}
\bibinfo{author}{\bibfnamefont{D.}~\bibnamefont{Poland}} \bibnamefont{and}
  \bibinfo{author}{\bibfnamefont{H.~A.} \bibnamefont{Scheraga}},
  \bibinfo{journal}{J. Chem. Phys.} \textbf{\bibinfo{volume}{45}},
  \bibinfo{pages}{1464} (\bibinfo{year}{1966}{\natexlab{b}}).

\bibitem[{\citenamefont{Cule and Hwa}(1997)}]{cule}
\bibinfo{author}{\bibfnamefont{D.}~\bibnamefont{Cule}} \bibnamefont{and}
  \bibinfo{author}{\bibfnamefont{T.}~\bibnamefont{Hwa}},
  \bibinfo{journal}{Physical Review Letters} \textbf{\bibinfo{volume}{79}},
  \bibinfo{pages}{2375} (\bibinfo{year}{1997}), \bibinfo{note}{article Aps
  Access restricted.}

\bibitem[{\citenamefont{Singh and Singh}(2001)}]{singh}
\bibinfo{author}{\bibfnamefont{N.}~\bibnamefont{Singh}} \bibnamefont{and}
  \bibinfo{author}{\bibfnamefont{Y.}~\bibnamefont{Singh}},
  \bibinfo{journal}{Physical Review E: Statistical, Nonlinear, and Soft Matter
  Physics} \textbf{\bibinfo{volume}{64}}, \bibinfo{pages}{042901/1}
  (\bibinfo{year}{2001}).

\bibitem[{\citenamefont{Kafri et~al.}(2000)\citenamefont{Kafri, Mukamel, and
  Peliti}}]{kafri}
\bibinfo{author}{\bibfnamefont{Y.}~\bibnamefont{Kafri}},
  \bibinfo{author}{\bibfnamefont{D.}~\bibnamefont{Mukamel}}, \bibnamefont{and}
  \bibinfo{author}{\bibfnamefont{L.}~\bibnamefont{Peliti}},
  \bibinfo{journal}{Physical Review Letters} \textbf{\bibinfo{volume}{85}},
  \bibinfo{pages}{4988} (\bibinfo{year}{2000}).

\bibitem[{\citenamefont{Peyrard and Bishop}(1989)}]{peyrard}
\bibinfo{author}{\bibfnamefont{M.}~\bibnamefont{Peyrard}} \bibnamefont{and}
  \bibinfo{author}{\bibfnamefont{A.~R.} \bibnamefont{Bishop}},
  \bibinfo{journal}{Physical Review Letters} \textbf{\bibinfo{volume}{62}},
  \bibinfo{pages}{2755} (\bibinfo{year}{1989}), \bibinfo{note}{article}.

\bibitem[{not()}]{note}
\bibinfo{note}{In its simplest form, this family of models takes only partial
  account of the effects of excluded volume and ignores the influences of
  base-pair inhomogeneity and torsional strain associated with the formation of
  denaturation bubbles. Much of the recent theoretical literature in DNA
  melting is concerned with attempts to take proper account of the above and
  other effects on DNA melting.}

\bibitem[{\citenamefont{Cantor and Schimmel}(1980{\natexlab{a}})}]{cantor1}
\bibinfo{author}{\bibfnamefont{C.~R.} \bibnamefont{Cantor}} \bibnamefont{and}
  \bibinfo{author}{\bibfnamefont{P.~R.} \bibnamefont{Schimmel}},
  \emph{\bibinfo{title}{Techniques for the study of biological structure and
  function}} (\bibinfo{publisher}{W. H. Freeman}, \bibinfo{address}{San
  Francisco}, \bibinfo{year}{1980}{\natexlab{a}}).

\bibitem[{\citenamefont{Kauffman et~al.}(2002)\citenamefont{Kauffman, Gruner,
  and Zocchi}}]{gruner}
\bibinfo{author}{\bibfnamefont{E.}~\bibnamefont{Kauffman}},
  \bibinfo{author}{\bibfnamefont{G.}~\bibnamefont{Gruner}}, \bibnamefont{and}
  \bibinfo{author}{\bibfnamefont{G.}~\bibnamefont{Zocchi}}
  (\bibinfo{year}{2002}).

\bibitem[{\citenamefont{Cantor and Schimmel}(1980{\natexlab{b}})}]{cantor2}
\bibinfo{author}{\bibfnamefont{C.~R.} \bibnamefont{Cantor}} \bibnamefont{and}
  \bibinfo{author}{\bibfnamefont{P.~R.} \bibnamefont{Schimmel}},
  \emph{\bibinfo{title}{The behavior of biological macromolecules}}
  (\bibinfo{publisher}{W. H. Freeman}, \bibinfo{address}{San Francisco},
  \bibinfo{year}{1980}{\natexlab{b}}).

\bibitem[{\citenamefont{Cantor and Schimmel}(1980{\natexlab{c}})}]{cantor3}
\bibinfo{author}{\bibfnamefont{C.~R.} \bibnamefont{Cantor}} \bibnamefont{and}
  \bibinfo{author}{\bibfnamefont{P.~R.} \bibnamefont{Schimmel}},
  \emph{\bibinfo{title}{The conformation of biological macromolecules}}
  (\bibinfo{publisher}{W. H. Freeman}, \bibinfo{address}{San Francisco},
  \bibinfo{year}{1980}{\natexlab{c}}), \bibinfo{note}{charles R. Cantor, Paul
  R. Schimmel. ill. ; 24 cm.}

\end{thebibliography}

\end{document}